\documentclass[sigconf,natbib=true,anonymous=false]{acmart}

\usepackage{microtype}
\usepackage{comment}
\usepackage{tcolorbox}
\usepackage{amsmath}
\usepackage{tabularray}
\usepackage{soul}
\usepackage{amsmath}
\usepackage{algorithm}
\usepackage[noend]{algpseudocode}

\newcommand{\sys}{\emph{Contriever4GR}}
\newcommand{\pt}{\textbf{pt}-\sys}
\newcommand{\ft}{\textbf{ft}-\sys}
\newcommand{\cpvanilla}{\textbf{ct$^v$}-\textbf{pt}-\sys}
\newcommand{\cpipt}{\textbf{ct$^i$}-\textbf{pt}-\sys}
\newcommand{\cpift}{\textbf{ct$^i$}-\textbf{ft}-\sys}
\newcommand{\cpspt}{\textbf{ct$^s$}-\textbf{pt}-\sys}
\newcommand{\cpsft}{\textbf{ct$^s$}-\textbf{ft}-\sys}

\DeclareMathOperator*{\argmax}{argmax}

\AtBeginDocument{%
  \providecommand\BibTeX{{%
    \normalfont B\kern-0.5em{\scshape i\kern-0.25em b}\kern-0.8em\TeX}}}

\copyrightyear{2024}
\acmYear{2024}
\setcopyright{acmlicensed}\acmConference[SIGIR '24]{Proceedings of the 47th International ACM SIGIR Conference on Research and Development in Information Retrieval}{July 14--18, 2024}{Washington, DC, USA}
\acmBooktitle{Proceedings of the 47th International ACM SIGIR Conference on Research and Development in Information Retrieval (SIGIR '24), July 14--18, 2024, Washington, DC, USA}
\acmDOI{10.1145/3626772.3657744}
\acmISBN{979-8-4007-0431-4/24/07}

\begin{document}

\title{Graded Relevance Scoring of Written Essays with Dense Retrieval}

\author{Salam Albatarni}
\email{sa1800633@qu.edu.qa}
\affiliation{%
  \institution{Qatar University}
  \city{Doha}
  \country{Qatar}
}

\author{Sohaila Eltanbouly}
\email{se1403101@qu.edu.qa}
\affiliation{%
  \institution{Qatar University}
  \city{Doha}
  \country{Qatar}
}

\author{Tamer Elsayed}
\email{telsayed@qu.edu.qa}
\affiliation{%
  \institution{Qatar University}
  \city{Doha}
  \country{Qatar}
}

\begin{abstract}

Automated Essay Scoring automates the grading process of essays, providing a great advantage for improving the writing proficiency of students. While holistic essay scoring research is prevalent, a noticeable gap exists in scoring essays for specific quality traits. 
In this work, we focus on the relevance trait, which measures the ability of the student to stay on-topic throughout the entire essay. 
We propose a novel approach for graded relevance scoring of written essays that employs dense retrieval encoders.
Dense representations of essays at different relevance levels then form clusters in the embeddings space, such that their centroids are potentially separate enough to effectively represent their relevance levels. We hence use the simple 1-Nearest-Neighbor classification over those centroids to determine the relevance level of an unseen essay. As an effective unsupervised dense encoder, we leverage Contriever, which is pre-trained with contrastive learning and demonstrated comparable performance to supervised dense retrieval models. 
We tested our approach on both task-specific (i.e., training and testing on same task) and cross-task (i.e., testing on unseen task) scenarios using the widely used ASAP++ dataset. 
Our method establishes a new state-of-the-art performance in the task-specific scenario, while its extension for the cross-task scenario exhibited a performance that is on par with the state-of-the-art model for that scenario. 
We also analyzed the performance of our approach in a more practical few-shot scenario, showing that it can significantly reduce the labeling cost while sacrificing only 10\% of its effectiveness.

\end{abstract}



\maketitle

\section{Introduction}
Automated Essay Scoring (AES) aims to automatically assign a quality score to written essays. It comes in handy for teachers, alleviating the burden of correcting numerous essays and allowing them to concentrate on more crucial responsibilities. Furthermore, AES has the advantage of providing quick and consistent feedback to students, improving the learning process.
AES research has primarily focused on holistic scoring \cite{ke2019automated}, which provides a single score reflecting the overall quality of the essay. However, in terms of practicality and effectiveness, a single score falls short in guiding students on how to enhance their skills. 
Trait-based AES \cite{mathias-bhattacharyya-2020-neural} fills this gap by individually scoring \emph{traits} of quality, e.g., organization, development, and relevance. Trait scoring enables students to gain insights into specific areas of improvement, empowering them to understand their weaknesses and enhance their writing proficiency. 

In this work, we focus on scoring the \emph{relevance} (or so-called \emph{prompt adherence}) trait. The relevance trait, in particular, evaluates the extent to which the essay aligns with the given \emph{task-prompt}.\footnote{The task-prompt refers to the specific instructions or guidelines provided to students to guide their essay writing on a particular \emph{topic}.} This trait is crucial as it gauges the student's ability to stay on topic, i.e., maintaining a clear and direct connection to the main subject throughout the writing. Several studies used the essence of prompt adherence as a means to aid in holistic scoring~\cite{chen2018relevance}; 
however, there is a lack of specific scoring approaches for the relevance trait. 

AES systems can be categorized into two types based on the training and inference settings. \emph{Task-specific} AES focuses on a single writing task, where the system is trained exclusively on essays written for \emph{one} specific task, then, during inference, it grades unseen essays written for the \emph{same task}.
This setup enables the AES system to learn the distinctive characteristics associated with a specific task, allowing for a more precise assessment of the essays written in response to that task. This type is predominant in the literature for both holistic and trait-based AES. The early research on task-specific AES used feature-based learning approaches~\cite{persing-modeling,mathias2018asap++}, then, 
different neural-based approaches have been proposed \cite{dong2017attention,mathias-bhattacharyya-2020-neural,ijcai2020p536}. Subsequently, 
pre-trained language models have become dominant for developing AES systems~\cite{yang-etal-2020-enhancing,kumar2021many,9530411,jiang-etal}. Most of those works focused on holistic scoring, and less attention has been given to scoring individual traits. 

Meanwhile, \emph{cross-task} (also known as \emph{cross-prompt}) AES systems are trained on essays written for multiple \emph{source} writing tasks to grade essays written for \emph{unseen target} tasks~\cite{jiang-etal}. In this setting, the AES system leverages insights gained from various writing tasks to score essays from unseen tasks. It is commonly observed in real-world scenarios that limited data is available for target tasks~\cite{jin-etal-2018-tdnn}, which emphasizes the necessity for developing generalizable cross-task AES systems for grading essays for unseen tasks. Multiple research studies have proposed cross-task holistic scoring systems, e.g., \cite{jin-etal-2018-tdnn,li2020sednn,cao2020domain}. More recently,  
other approaches for cross-task AES systems have been proposed to score individual traits of the essay along with the holistic score, e.g., \cite{ridley2021automated,do-etal-2023-prompt,chen-li-2023-pmaes}. Nevertheless, all traits are handled the same, without employing a distinct approach for any specific trait, overlooking the fact that each trait focuses on a different dimension of the writing quality of the essay.

To that end, in this paper, we focus on scoring the relevance trait for both \emph{task-specific} and \emph{cross-task} scenarios. 
We propose a \emph{novel} approach for \emph{graded relevance} scoring (i.e., scoring into one of multiple relevance grades or levels) of written essays. It employs dense retrieval encoders to represent training essays in the embedding space. We then hypothesize that the dense representations of essays having the same relevance level form a cluster in that space such that the centroids of the clusters of different relevance levels will be separate enough to effectively represent their respective relevance levels.
We hence use the simple $1NN$ classification model over those centroids to determine the relevance level of an unseen essay. 
As an example of effective unsupervised dense encoder, we leverage Contriever~\cite{izacard2021towards}, which is pre-trained with contrastive learning and has demonstrated comparable performance to supervised dense retrieval models. 

In the context of graded relevance scoring of written essays, we address the following research questions:

\begin{itemize}
    \item \textbf{RQ1}: Can the \emph{pre-trained} Contriever model be effectively leveraged for the task? (\textbf{Out-of-the-box scenario})
    \item \textbf{RQ2}: What is the impact of \emph{fine-tuning} the Contriever model on the performance? (\textbf{Fine-tuning scenario})
    \item \textbf{RQ3}: In case only \emph{few} training essays are available for each relevance level, how would that affect the performance? (\textbf{Few-shot learning scenario})
    \item \textbf{RQ4}: In case \emph{no} training essays are available, can we effectively leverage previously-labeled essays from other tasks? (\textbf{Cross-task scenario}) 
\end{itemize}

Our study yields promising results when scoring the relevance trait using the pre-trained Contriever. However, fine-tuning Contriever outperformed the state-of-the-art (SOTA) models. Furthermore, we propose an extension to our approach for the cross-task setup. We transform the approach to \emph{task-independent} by ``excluding'' the information of the task-prompt from the essays. This simple trick resulted in a performance boost, achieving an on par performance with the SOTA model for that scenario. 

In summary, our main contribution is four-fold:
 \begin{itemize}
    \item We propose a novel approach that employs dense retrieval for the general task of graded relevance scoring.
    \item Our approach establishes a new SOTA performance on scoring the relevance of essays in the task-specific setup.
    \item We analyze the performance of our approach in a more practical few-shot scenario, showing a significant saving of labeling cost while sacrificing only 10\% of the effectiveness.
    \item We propose a simple but effective cross-task extension of our approach that made it on par with the SOTA performance.
 \end{itemize}

The rest of the paper is organized as follows. Section \ref{sec:related-work} discusses the work related to relevance trait scoring for task-specific and cross-task setups. A detailed description of our approach is provided in Section \ref{sec:methodology}. Section \ref{sec:experimental_setup} presents the experimental setup. Section \ref{sec:experimental_evaluation} answers our research questions. Finally, we conclude in Section \ref{sec:conclusion}.

\section{Related Work} \label{sec:related-work}

In this section, we offer a comprehensive review of existing methods for task-specific and cross-task techniques for relevance scoring. We address the limitations of prior work and highlight the unique contributions and distinctions of our approach.

\paragraph{\textbf{Task-specific Scoring}}

When it comes to AES, task-specific holistic scoring is predominantly emphasized, with less attention directed towards trait scoring. 
The early research concerning the scoring of the relevance trait focused on feature-based approaches. \citet{persing-modeling} pioneered the focused exploration of the relevance trait. They used a linear SVM regression model with a rich set of lexical and knowledge-based features to measure the relevance between the essays and the task-prompt. \citet{mathias2018asap++} used a common feature set with a Random Forest (RF) classifier for predicting the holistic and traits scores. 

Others employed traditional retrieval approaches, such as TF-IDF and pseudo-relevance feedback (PRF), to measure the similarity between the essay and the task-prompt. \citet{cummins2016unsupervised} expanded the task-prompt, then the relevance score was computed as the cosine similarity between the TF-IDF representations of the essay and the expanded prompt. The expansion terms were selected based on the closest words to the prompt vector, constructed using random-indexing, CBoW, skip-gram, and PRF. Similarly, \cite{rei-cummins-2016-sentence} used TF-IDF, CBoW, and skip-thoughts models to measure the similarity between the prompt and each sentence in the essay. \citet{chen2018relevance} used the relevance of the essay to the prompt as a feature for holistic scoring. The essay and prompt representations were acquired through an attention-based RNN, with the relevance score being computed via element-wise multiplication between the essay and prompt vectors. 

In light of the limited research on trait scoring, the objective of \citet{mathias-bhattacharyya-2020-neural} was to leverage various models originally intended for holistic scoring to score multiple essay traits, including relevance. Three approaches were used: a feature-based model with RF algorithm \cite{mathias2018asap++}, a string kernel-base approach \cite{cozma-etal-2018-automated}, and an attention-based neural model \cite{dong2017attention}. 
Most recently, \citet{kumar2021many} used a multi-task neural model to predict multiple writing trait scores in parallel (auxiliary tasks), and these scores were used by the network to predict the holistic score (primary task). They also tried to set one trait as a primary task and other traits along with the holistic score as auxiliary tasks. 

Although there are some approaches that have targeted the relevance trait, the majority of those approaches rely on feature-based and traditional retrieval methods. Moreover, none of the recent approaches focused on the relevance trait in specific; rather, they offered a generalized model for all traits, ignoring the fact that each trait pertains to different aspects within the essay. To the best of our knowledge, this is the first work that leverages the recent advancement in the dense information retrieval domain for the task of AES for the relevance trait. 

\paragraph{\textbf{Cross-task Scoring}}
Cross-task AES aims to train a model using labeled essays from one or more source tasks and then apply the model to score essays from a target task. Similar to task-specific research, predominant studies on cross-task evaluation have focused on holistic scoring \cite{jin-etal-2018-tdnn,li2020sednn,ridley2020prompt}. The cross-task trait-based AES was introduced by \citet{ridley2021automated}, where all of the efforts previously focused solely on holistic scoring. Their approach is an extension of the work of \citet{ridley2020prompt}, which utilized part-of-speech embedding with a convolution network to generate the essay’s representation. The architecture was modified by introducing shared low-level layers to learn common representation across tasks and high-level layers to capture task-specific information. ProTACT model \cite{do-etal-2023-prompt} also employed the idea of hierarchical representation, employing low-level layers for information sharing across traits and top layers for trait-specific information. Moreover, the task-prompt was utilized to acquire prompt-aware representation by applying essay-prompt attention. \citet{chen-li-2023-pmaes} proposed PMAES, a framework designed to improve cross-task representation through a prompt-mapping contrastive learning strategy. This approach involved the projection of source task essays onto target tasks, generating mapping representations specific to the target task. The objective was to minimize the distance between these mapping pairs, thereby aligning source and target tasks to achieve greater consistency in their representations.

Trait cross-task AES has received less attention compared to task-specific AES. Similar to the task-specific approaches, all of the existing solutions for cross-task provided a common framework for all the traits. Our cross-task approach is tailored specifically for the relevance trait. This is done by generating task-independent representations, which effectively map essays with similar relevance levels across different tasks closer together. One of the unique aspects of our approach is its adaptability for both task-specific and cross-task scenarios, requiring only minor modifications.

\section{Proposed Approach} \label{sec:methodology}
\begin{figure*}[h]
    \centering
    \includegraphics[scale=0.6]{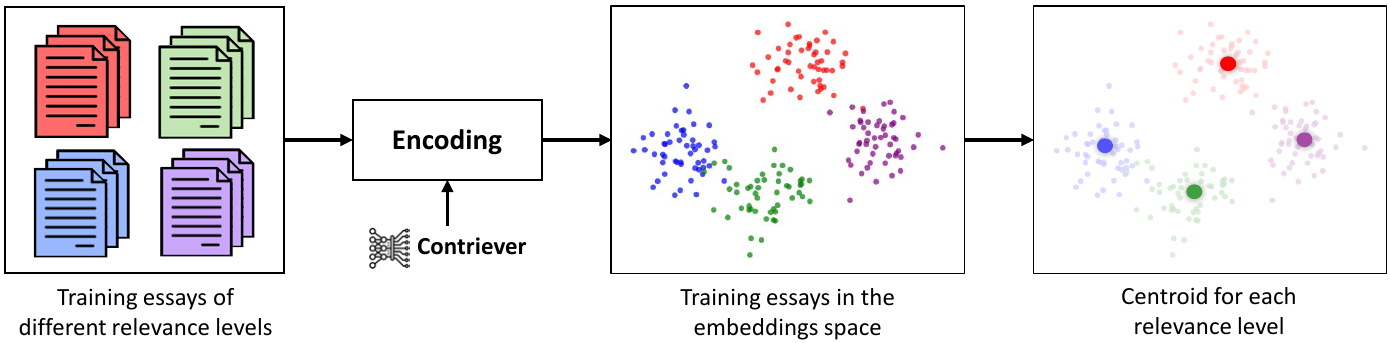}
    \caption{Training phase of our proposed approach for graded relevance.}
    \label{fig:ts-approach}
\end{figure*}

In this section, we present and discuss in detail our proposed approach for graded relevance scoring of essays in different scenarios. We address the task-specific scenario in Section \ref{sec:contriever-method}, the fine-tuning scenario in Section \ref{sec:contriever-finetuning}, and the cross-task scenario in Section \ref{sec:contriever-crossprompt}.

\subsection{Graded Relevance Scoring of Essays with Dense Retrieval}\label{sec:contriever-method}

Dense retrieval models are generally encoders that are trained to generate dense vector representations of documents (and queries) so that texts that are topically similar are close in the embedding space, while those that are topically distant are further apart in that space. This can then be employed to find documents that are relevant to a given query, in a typical retrieval scenario, by computing the similarity between the dense representations of the query on one side and documents on the other side. 

Our approach builds on this idea to score the relevance trait of students' written essays given a task-prompt, but from a different angle. While the obvious approach is to use the prompt as a query and the test essays as documents in the above setup, several challenges arise. Notably, prompts often differ in length from essays, making the mapping of similarity scores to relevance levels harder. Furthermore, this approach may not capture the diverse writing proficiency levels exhibited in essays, particularly those written by school students. This variability adds complexity to the problem. 

Our approach relies heavily on \emph{two} main components: having examples of labeled (i.e., manually-graded or scored) essays from \emph{each} relevance level, and utilizing a robust dense encoder. We hypothesize that the encoder (i.e., retrieval model) can effectively encode essays in a manner that positions those from the same relevance level in close proximity while keeping essays from different relevance levels distant. This yields a (somewhat) separate cluster of essays (in the embedding space) for each different relevance level. Therefore, \emph{centroids} of those clusters can serve as reference points (or good representatives) of their corresponding relevance levels. 

This constitutes the ``training'' phase of our approach. 
During inference (when we get unseen essays to score), we simply classify essays by assigning them the relevance level corresponding to the \emph{nearest} centroid. 
In other words, our approach employs a \emph{1-Nearest-Neighbor} (\emph{1NN}) algorithm over the centroids corresponding to the different relevance levels.
 
This innovative setup reframes the problem; it treats the test essay (i.e., the one to be scored) as a query and the centroids as documents. Our goal is to identify the most ``relevant'' document (centroid) for the given query (test essay), effectively assigning a relevance score based on this proximity in the embedding space. This approach also transforms the problem into a multi-class classification scenario, where the classes represent the various relevance levels.
It is worth mentioning that the model built here is a \emph{task-specific} model, as it is trained on essays labeled for a given task-prompt. Figure \ref{fig:ts-approach} illustrates the training phase of our approach.


Formally, we are given a task-prompt $p$ and a corresponding training set of essays $S_p$. Each training essay $s \in S_p$ is assigned a relevance level $R(s)$. We denote the set of all training essays that are assigned the relevance level $i$ as $S_p^i$, where $S_p^i = \{s \in S_p | R(s) = i\}$. 

We first encode each training essay $s$ using the dense encoder $D$ to get its dense vector representation $\vec{e}_s$, where $\vec{e}_s = D(s)$. 

Next, for each relevance level $i$, we compute its corresponding centroid $\vec{C}_i$ as the mean of the essay vectors of that level, as follows.
\begin{equation}
    \vec{C}_i = \frac{1}{|S_p^i|} \sum_{s \in S_p^i}\vec{e}_s
\end{equation}
\noindent where $|S_p^i|$ is the number of training essays that are assigned the relevance level $i$. 

Subsequently, for a given test essay $t$, we estimate its relevance level $R(t)$ based on its most similar relevance centroid, as follows.
\begin{equation} \label{test_essay_eq}
  R(t) = \argmax_i \phi(\vec{e_t}, \vec{C}_i)
\end{equation}
\noindent where $\phi$ represents the similarity function used to measure the closeness of the test essay and the relevance centroids.

As mentioned earlier, our method relies heavily on the effectiveness of the encoder model used to represent the essays. Dense retrieval models are originally trained for relevance tasks, thus are perfect option to serve our purpose in scoring the \emph{relevance} trait. 

As there are several dense retrieval models in the literature, we opted to choose Contriever \cite{izacard2021towards}, an unsupervised dense retrieval model trained with contrastive learning. 
There are several reasons why we believe Contriever fits our needs. 
Firstly, it is trained with \emph{contrastive} learning focusing exclusively on distinguishing between similar and non-similar pairs of texts, which is highly needed in our problem. Moreover, it is pre-trained with a large amount of data, and being an unsupervised model makes it less susceptible to bias toward a specific topic or domain. Furthermore, it showed out-of-the-box strong performance in earlier retrieval tasks \cite{gao-etal-2023-precise}.

Notice that the Contriever encoder is used thus far in its original \emph{pre-trained} form. We therefore call this scenario the \emph{out-of-the-box} scenario. We also denote the overall approach we introduced here by \pt, the \textbf{pre-trained} Contriever for Graded Relevance, to distinguish it from the \emph{fine-tuned} and \emph{cross-task} models we introduce in the subsequent sections.

\subsection{Fine-tuning for Task-specific Scoring}\label{sec:contriever-finetuning}
A natural decision is to fine-tune pre-trained models for downstream tasks. Fine-tuning allows the model to learn more information about a specific task and, hence, potentially achieve better performance on that task. As fine-tuning dense retrieval models is confined to retrieval tasks, we propose a slightly different setup for fine-tuning the retrieval model for graded relevance. 

In tasks involving retrieval with dense models, the training dataset comprises queries, relevant, and non-relevant documents. The primary objective is to train the model to decide whether a given document is relevant to a specific query. In our context, essays within the same relevance level are considered ``relevant to each other,'' and essays with different relevance levels are considered ``non-relevant to each other.'' Hence, our model needs to be optimized to differentiate between the different relevance levels. 

In Contriever, the contrastive learning setup requires triplets within the training dataset. These triplets comprise anchor, positive, and negative examples. Positive instances are derived through random cropping from the anchor example, while negative examples are sampled from the batch using MoCo \cite{izacard2021towards}. In our scenario, the process is more straightforward. Positive examples consist of essays within the same relevance level as the anchor essay, while negative examples encompass essays within other relevance levels. 

Contriever model was initially trained with InfoNCE loss, presented in Equation \ref{equ:infonce}. InfoNCE loss is a categorical cross-entropy loss over the similarity between the anchor and positive and the anchor and all negative examples \cite{oord2018representation}, and it optimizes the negative log probability of classifying the positive sample correctly.

\begin{equation}\label{equ:infonce}
    \mathcal{L}_\text{InfoNCE}(s_a , s^+, N) = -\log{\frac{e^{\phi(s_a, s^+)/\tau}}{e^{\phi(s_a, s^+)/\tau}+\sum_{s^- \in N}e^{\phi(s_a, s^-)/\tau}}}
\end{equation}
\noindent where $s_a$, $s^+$, $s^-$, and $N$ denote the encoded anchor, positive, negative, and set of negative essays, respectively, and $\tau$ is a temperature parameter.

Pairwise softmax cross-entropy (PSCE) loss is another loss function, used in fine-tuning another dense retrieval model~\cite{khattab2020colbert}, that serves the same objective. It incorporates the similarity between positive and negative examples with the anchor example. 
\begin{equation} \label{equ:loss}
    \mathcal{L}_\text{PSCE}(s_a , s^+, s^-)=-\log{\frac{e^{\phi({s_a , s^+})}}{e^{\phi({s_a , s^+})}+e^{\phi({s_a , s^-})}}}
\end{equation}
\noindent where $s_a$, $s^+$, and $s^-$ denote the encoded anchor, positive, and negative essays, respectively, constituting a training triplet.

This setup raises multiple alternatives in terms of how the negative training examples are sampled. For an anchor essay example $s_a$ of relevance level $R(s_a)$, we can sample negative examples $s^-$ from all the other relevance levels, i.e., $R(s_a) \neq R(s^-)$. We denote this as sampling from \emph{All} levels. This exposes the model to training triplets from all different levels.
We can also restrict the sampling to the relevance levels that are relatively far from the relevance level of the anchor, i.e., $R(s_a) \neq R(s^-)\pm 1$. We denote this as sampling from \emph{Easy} levels, since they should be easy to distinguish. The rationale is that essays with closer scores might actually have the same score if rated by a different rater, potentially confusing the model.
Alternatively, we can restrict the sampling to the relevance levels that are closest to the relevance level of the anchor, i.e., $R(s_a) = R(s^-)\pm 1$. We denote this as sampling from \emph{Hard} levels, since they should be hard to distinguish. 
The rationale is that by mastering this, the model will also be better at distinguishing relevance levels with larger margins.
In each case, the number of negative examples we sample from each selected level might vary.

As the Contriever encoder here is fine-tuned, we, hereafter, denote such model by \ft, the \textbf{fine-tuned} Contriever for Graded Relevance.

\subsection{Cross-Task Scoring}\label{sec:contriever-crossprompt}

\begin{figure*}[h]
    \centering
\includegraphics[width=\textwidth]{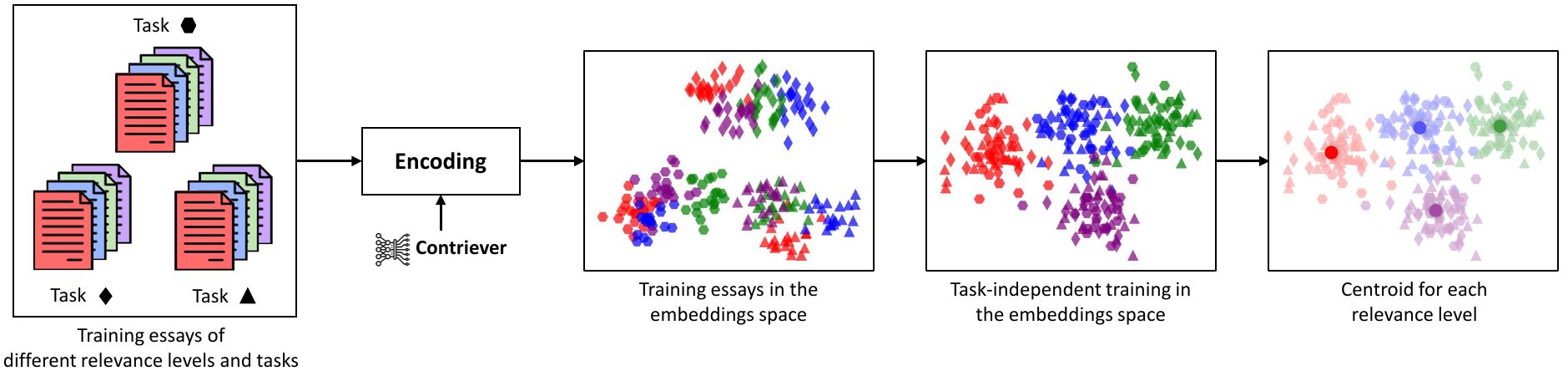}
    \caption{Training phase of our proposed cross-task approach with task-independent representations.}
    \label{fig:cross-prompt}
\end{figure*}

In the earlier scenarios, we assume there are labeled (i.e., scored or graded) essays for the given writing task, which are needed to build a task-specific model, whether it is pre-trained or fine-tuned. 
However, this is not always available. We might need to build a model with \emph{no} labeled essays for the test task, i.e., a ``zero-shot'' model. In such a case, the only available labeled essays to learn from come from \emph{other} different tasks than the test task, hence denoted as the \emph{cross-task} scenario. In the cross-task setup, the training typically utilizes multiple \emph{source} tasks, whereas testing is conducted on an \emph{unseen target} task. 

Applying the pre-trained or fine-tuned \sys{} approach directly to the cross-task scenario (by just training on the superset of all the training essays of the source tasks) would yield separate clusters (in the embedding space) of task-related essays rather than clusters of essays of same relevance level regardless of the task.
Hence, the challenge of that scenario in our approach is how to ``normalize'' the representation of the essays in order to get ``task-independent'' representations. 

Since the dense encoder represents texts in an embedding ``semantic'' space, it is tempting to make those representations \emph{task-independent} by simply subtracting the task-prompt representation, disentangling the topic-specific semantic features from the relevance level features. 
Consequently, the cluster of each relevance level will contain the normalized task-independent essay embeddings from all the source tasks. Accordingly, the computed centroids are potentially task-independent. Similar to the earlier scenario, $1NN$ is used to classify the normalized essays for the target task. 

Formally, we are given a set of source tasks $P$, each with a task-prompt $p_n$ and a corresponding set of labeled essays $S_{p_n}$. For a relevance level $i$, the corresponding centroid $\vec{C}_i$ is computed as:

\begin{equation}
    \vec{C}_i = \frac{1}{\sum_{p_n \in P}|S_{p_n}^i|} \sum_{p_n \in P} \sum_{s \in S_{p_n}^i} \vec{e}_s - \vec{p}_n
\end{equation}
\noindent The relevance level $R(t)$ for a test essay $t$ is then computed 
using Equation \ref{test_essay_eq} except that $\vec{e}_t$ is now normalized as $\vec{e}_t - \vec{p}_t$. 

This approach scores the essay based on the aggregated information of the relevance levels from the source tasks. Nevertheless, an essential piece of information that can also be considered is the similarity with the target task-prompt. Consequently, we employed both the scores determined by the nearest centroid and the similarity with the target task-prompt to assign the final relevance score of the essay. 
For a test essay $t$ and a target task-prompt $p^{\prime}$ that has a maximum relevance score of $r_{\max}$, the similarity score $S(t)$ is normalized as follows: 

\begin{equation} 
    S(t) = \phi(\vec{e}_t,\vec{p}^{\prime}) * r_{\max}
\end{equation}

\noindent where $\phi$ is the similarity function with a [0-1] output range. Then, the final score $R^{*}(t)$ of the test essay $t$ is computed as the average between the two scores:

\begin{equation}
    R^{*}(t) = \frac{1}{2} (R(t) + S(t))
\end{equation}

The fine-tuning process for Contriever in the cross-task scenario follows the same setup for the task-specific scenario. The only difference lies in the essays' representations, which are made task-independent by subtracting the embedding vector of the task-prompt from the anchor, positive, and negative examples during the fine-tuning process. 

Our general approach for the cross-task scenario is hereafter denoted by a prefix \textbf{ct} amended to the pre-trained or fine-tuned model names for Graded Relevance.

\section{Experimental Setup} \label{sec:experimental_setup}

In this section, we introduce the setup used to conduct our experiments. We first present the dataset we used for evaluation and the specific evaluation measure for the AES task. We next discuss the hyper-parameters we used for fine-tuning, before we list the baselines with which we compare our work for the task-specific and cross-task approaches.
 
\label{sec:dataset}
\noindent\paragraph{\textbf{Dataset}} We employed the commonly-used Automated Student’s Assessment Prize (ASAP)\footnote{\url{https://www.kaggle.com/c/asap-aes}} and ASAP++ \cite{mathias2018asap++} datasets. ASAP dataset comprises 8 different tasks $T$, where each task has a set of written essays. The original ASAP dataset contains the holistic scores for all tasks and the trait scores only for T7 and T8. ASAP++ extends ASAP by scoring the essay traits for tasks T1-T6. 
To test our approach, we used the tasks that have annotations for the relevance trait, namely, T3, T4, T5, and T6.
Table~\ref{tab:dataset} summarizes the dataset we used in our experiment. Following previous work, we use 5-fold cross-validation with the same folds used by \citet{taghipourng2016neural} for task-specific models. 

\begin{table}[h]
\centering
\caption{ASAP++ relevance trait tasks.}
\label{tab:dataset}
\begin{tblr}{
  cells = {c},
  hline{1-2,6} = {-}{},
  rowsep=1pt, 
}
Task & Relevance Levels & Ave. Length & Essays \\
T3    & 0-3        &  100       & 1726         \\
T4    & 0-3        &  100       &   1772        \\
T5    & 0-4        &  125       &  1805         \\
T6    & 0-4       &  150       &   1800         
\end{tblr}
\end{table}

\label{sec:measure}
\noindent\paragraph{\textbf{Evaluation Measure}} For evaluating our approach, we used Quadratic Weighted Kappa (QWK) \cite{cohen1968weighted}, which is a widely used measure for AES. QWK measures the agreement between the scores of two raters, in our case, the human rater and the system. QWK is suitable for this task as it weighs the degree of disagreement between the raters, which is what we want as the scores are \emph{ordered}.

\label{sec:hyperparameters}
\noindent\paragraph{\textbf{Implementation and Hyper-parameters}} All of our experiments are carried out with the Pytorch library.\footnote{\url{https://pytorch.org/}} For Contriever, we used the available checkpoint accessible on Hugging Face's model hub.\footnote{\url{https://huggingface.co/facebook/contriever}}
For the hyper-parameter tuning, we employed a dynamic learning rate scheduler, ReduceLROnPlateau, which adjusts the learning rate by a factor of 0.5 upon observing a decrease in QWK on the validation set, with patience of 2 epochs. We used the AdamW optimizer with an initial learning rate of 1e-6 and a batch size of 16. Moreover, to avoid overfitting, we used an early stopping condition to stop the training when the QWK on the validation set shows no improvement for 20 epochs. Then, we used the best model on the validation set for testing.

\label{sec:baselines}
\paragraph{\textbf{Task-specific Baselines}} We employed the following baselines:
\begin{itemize}
     \item \emph{Feature-based}: \citet{mathias-bhattacharyya-2020-neural} utilized multiple models initially designed for holistic scoring. One of their baselines is a feature-based model described in \cite{mathias2018asap++}.
    \item \emph{Attention-based Neural Model} \cite{mathias-bhattacharyya-2020-neural}: This model, which is based the work of~\citet{dong2017attention}, achieved the best results for the relevance trait. Hence, we use it as the SOTA baseline. 
    \item \emph{Multi-task Neural Model} \cite{kumar2021many}: This work developed a multi-task setup for holistic scoring by scoring the traits as sub-tasks. In a specific experiment, individual traits were set as the primary task for prediction, utilizing the other traits alongside the holistic score as sub-tasks. The reported results provided an average score across all tasks. As such, our comparison will be based on this averaged metric.
\end{itemize}

\paragraph{\textbf{Cross-task Baselines}} We employed the following baselines:
\begin{itemize}
    \item \emph{Vanilla Baseline}: To test the idea of removing topic information by subtracting the encoded task-prompt from the encoded essays, we developed a baseline model that uses the same approach for cross-task setup but without removing the topic information from the essays' representations. We denote this baseline as the ``vanilla'' cross-task approach \cpvanilla. 
    \item \emph{PMAES} \cite{chen-li-2023-pmaes}: This recent work developed a prompt-mapping contrastive learning strategy to capture the shared information between the source and target tasks. To score the traits, a different output layer is used for each trait. 
    \item \emph{ProTACT} \cite{do-etal-2023-prompt}: This work proposed a shared model to learn the prompt-aware essay representations. Subsequently, trait-specific layers were introduced on top of the shared layers to score the individual traits. ProTACT achieved the best performance for the relevance trait in the cross-task setup. Therefore, we use it as the SOTA baseline.
\end{itemize}

\section{Experimental Evaluation} \label{sec:experimental_evaluation}
In this section, we present and discuss the results of our experiments addressing the four research questions. Section \ref{sec:RQ1} illustrates the performance of the \pt{} model comparing it with SOTA baselines. Section \ref{sec:RQ2} details the fine-tuning process and results. Section \ref{sec:RQ3} discusses the performance of \pt{} and \ft{} in few-shot settings. Finally, Section \ref{sec:RQ4} offers a comprehensive analysis of the cross-task scenario.

\subsection{Pre-trained Contriever (RQ1)}  
\label{sec:RQ1}
We first examine the performance of our pre-trained model \pt, leveraging the Contriever model without any further training. Table \ref{tab:pre-trained-results} compares our model against the baseline models.

\begin{table}
\centering
\caption{Performance (in QWK) of our pre-trained \sys{} on the \underline{test} sets, compared to the baselines.}
\label{tab:pre-trained-results}
\resizebox{\linewidth}{!}{%
\begin{tblr}{
  cells = {c},
  hline{1-2,5-6} = {-}{},
  rowsep=1pt, 
}
\textbf{Model}& \textbf{T3}& \textbf{T4}& \textbf{T5}& \textbf{T6}& \textbf{Ave.}\\
Feature-based & 0.575 & 0.636 & 0.639 & 0.581 & 0.608          \\
Multi-task NM & -     & -     & -     & -     & 0.730          \\
Attn-based NM & 0.683 & 0.738 & 0.719 & 0.783 & \textbf{0.731} \\
\pt            & 0.633 & 0.673 & 0.633 & 0.693 & 0.658          
\end{tblr}
}
\end{table}

Remarkably, this simple and straightforward idea achieved an average QWK of 0.658 with an \emph{out-of-the-box} Contriever model, which is quite effective for the AES task. It even performs consistently across tasks. Furthermore, it clearly outperforms the feature-based baseline, with only about 7-point lag behind the SOTA model. This performance showcases the versatility of the Contriever model while underscoring the potential of our proposed framework.

\begin{table*}
\centering
\caption{Performance (in QWK) of different variations of \ft{} on the \underline{development} sets.}
\label{tab:val}
\begin{tblr}{
  row{even} = {c},
  row{1} = {c},
  row{3} = {c},
  row{5} = {c},
  row{9} = {c},
  row{11} = {c},
  row{13} = {c},
  cell{1}{1} = {c=5}{},
  cell{1}{6} = {c=5}{},
  cell{3}{1} = {r=2}{},
  cell{5}{1} = {r=3}{},
  cell{7}{2} = {c},
  cell{7}{3} = {c},
  cell{7}{4} = {c},
  cell{7}{5} = {c},
  cell{7}{9} = {c},
  cell{7}{10} = {c},
  cell{8}{1} = {r=2}{},
  cell{10}{1} = {r=4}{},
  vline{2} = {1}{},
  vline{6,10} = {2-13}{},
  hline{1-3,5,8,10,14} = {-}{},
  rowsep=1pt, 
}
\textbf{Configuration} &                         &                      &                         &                                                & \textbf{\textbf{Performance}} &                &                &                &                \\
\textbf{Exp.}          & \textbf{Similarity fn.} & \textbf{Loss fn.}    & {\textbf{Neg. Sampling}} & {\textbf{Neg. Samples/level}} & \textbf{T3}                   & \textbf{T4}    & \textbf{T5}    & \textbf{T6}    & \textbf{Ave.}  \\
\textbf{$A$}           & Cosine                  & PSCE        & All levels              & 1                                              & 0.718                         & 0.763          & 0.732          & 0.789          & 0.751          \\
                       & Euclidean               & PSCE        & All levels              & 1                                              & 0.710                         & 0.715          & 0.798          & 0.727          & 0.737          \\
\textbf{$B$}           & Cosine                  & InfoNCE ($\tau$=0.1) & All levels              & 1                                              & 0.658                         & 0.693          & 0.710          &  0.668             &   0.683            \\
                       & Cosine                  & InfoNCE ($\tau$=0.2) & All levels              & 1                                              & 0.648                         & 0.686          & 0.711          & 0.672             & 0.679              \\
                       & Cosine                  & InfoNCE ($\tau$=0.3) & All levels              & 1                                              & 0.662                         & 0.687          & 0.699          & 0.667              & 0.679              \\
\textbf{$C$}           & Cosine                  & PSCE        & Easy levels             & 1                                              & 0.675                         & 0.727          & 0.788          & 0.724          & 0.728          \\
                       & Cosine                  & PSCE        & Hard levels             & 1                                              & 0.658                         & 0.683          & 0.665          & 0.659          & 0.666          \\
\textbf{$D$}           & Cosine                  & PSCE        & All levels              & 2                                              & 0.720                         & 0.760          & 0.792          & 0.738          & 0.752          \\
                       & Cosine                  & PSCE        & All levels              & 3                                              & \textbf{0.726}                & 0.766          & \textbf{0.801} & 0.737          & 0.758          \\
                       & Cosine                  & PSCE        & All levels              & 4                                              & 0.722                         & \textbf{0.775} & 0.797          & 0.738          & 0.758          \\
                       & Cosine                  & PSCE        & All levels              & 5                                              & \textbf{0.726}                & 0.774          & 0.739          & \textbf{0.801} & \textbf{0.760} 
\end{tblr}
\end{table*}

\subsection{Effect of Fine-tuning (RQ2)} \label{sec:RQ2}
With the encouraging performance of the pre-trained model shown in the previous experiment, we then turn to examining the performance of our model after fine-tuning.

For fine-tuning our model, we have 4 hyper-parameters: the similarity function, the loss function, the selection criteria of negative training samples, and the number of negative training samples per relevance level. As it is infeasible to tune those hyper-parameters together with a grid search, we tune one at a time while fixing the others in a sequential series of experiments $A$ through $D$.
All experiments with different variations are conducted on the development sets. After tuning all parameters, we evaluate the best configuration on the test sets. Table \ref{tab:val} summarizes the performance of the tuning experiments on the development sets.

Firstly, we tested two similarity functions, cosine similarity and Euclidean distance, while using the PSCE loss function, selecting negative samples from all relevance levels, and choosing only 1 negative sample per relevance level. Cosine similarity is a metric frequently utilized for measuring the similarity between two vectors irrespective of their magnitudes. In contrast, Euclidean distance calculates the straight-line distance between two points, offering a different perspective on the similarity. The results of experiment $A$ show the superiority of the cosine similarity function, so we use it in the rest of the experiments.

For the next experiment $B$, we test the performance with the InfoNCE loss function, used for training Contriever \cite{izacard2021towards}, with different temperature values.
Surprisingly, although Contriever was pre-trained with InfoNCE loss, it performed less effectively compared to the PSCE loss for our task. This discrepancy may be closely tied to the nature of our problem reformulation. The later loss function has shown effectiveness for retrieval tasks, where the goal is to rank relevant documents higher than non-relevant ones. This aligns better with our problem setup. It is also worth noting that the use of InfoNCE in our setup might not have been optimal, where we have a limited number of negative samples per anchor essay. Accordingly, we continued our experiments with the PSCE loss. 

We then examine, in experiment $C$, the effect of the other ways of selecting negative samples (Easy and Hard levels), which differ in what relevance levels to choose the samples from. The results revealed that using Easy negative samples exhibited better performance, as the Contriever model was better able to learn from them to distinguish between the different relevance grades; however, combining both Easy and Hard levels (which comprise the All levels option, shown in experiment $A$) was even better, indicating that they are complementary and both are needed to learn a better model, and highlighting the benefit of providing the model with contrastive examples from various levels.

The final experiment $D$ examines the effect of training with different numbers of negative samples per relevance level. As expected, the more negative samples we include, the better the performance. However, the performance was slightly improving from 0.751 (with one sample per level) to 0.76 (with 5 samples per level). While this points to the importance of increasing the size of our training set, increasing it further incurs further cost in terms of training time and computing resources. 

Given the above findings, we settle on the following configuration for the final \ft{} model: cosine similarity as the similarity function, PSCE as the loss function, selecting negative samples from all relevance levels, and drawing 5 negative samples per each level. Table \ref{tab:fine-tuned-results} shows the performance of \ft{} on the \emph{test} sets, and, again, we contrast it with the baselines. In fact, \ft{} outperformed the SOTA model on 3 tasks out of 4, while improving the average score of all tasks with nearly 1 point, establishing a new SOTA for the problem of graded relevance scoring of written essays. Moreover, the results indicate that a simple use of a contrastively-learned dense retrieval model can match (and even outperform) the current SOTA models. 
These findings encourage utilizing dense retrieval models for downstream non-retrieval tasks, wherein a simple problem reformulation with 1NN can achieve impressive performance. 

Our best experimented setup is constrained by allowing only up to 5 negative samples per relevance level per anchor essay. However, it is worth noting that we have not explored the optimal scenario of including all possible negative examples for each anchor essay. We argue that doing so could potentially achieve even greater performance improvements.

\begin{table}[h]
\centering
\caption{Performance (in QWK) of \ft{} on the \underline{test} sets, compared to the baselines.}
\label{tab:fine-tuned-results}
\begin{tblr}{
  cells = {c},
  hline{1-2,5-6} = {-}{},
  rowsep=1pt, 
}
\textbf{Model}& \textbf{T3}& \textbf{T4}& \textbf{T5}& \textbf{T6}& \textbf{Ave.}\\
Feature-based \cite{mathias-bhattacharyya-2020-neural}                    & 0.575          & 0.636          & 0.639 & 0.581          & 0.608          \\
Multi-task NM \cite{kumar2021many} & - & - & - & - & 0.730 \\
Attn-based NM \cite{mathias-bhattacharyya-2020-neural}                   & 0.683          & 0.738          & \textbf{0.719} & 0.783          & 0.731          \\
\ft & \textbf{0.704} & \textbf{0.766} & 0.707 & \textbf{0.785} & \textbf{0.740} 
\end{tblr}
\end{table}

\begin{figure*}[h]
    \centering
    \includegraphics[width=\textwidth]{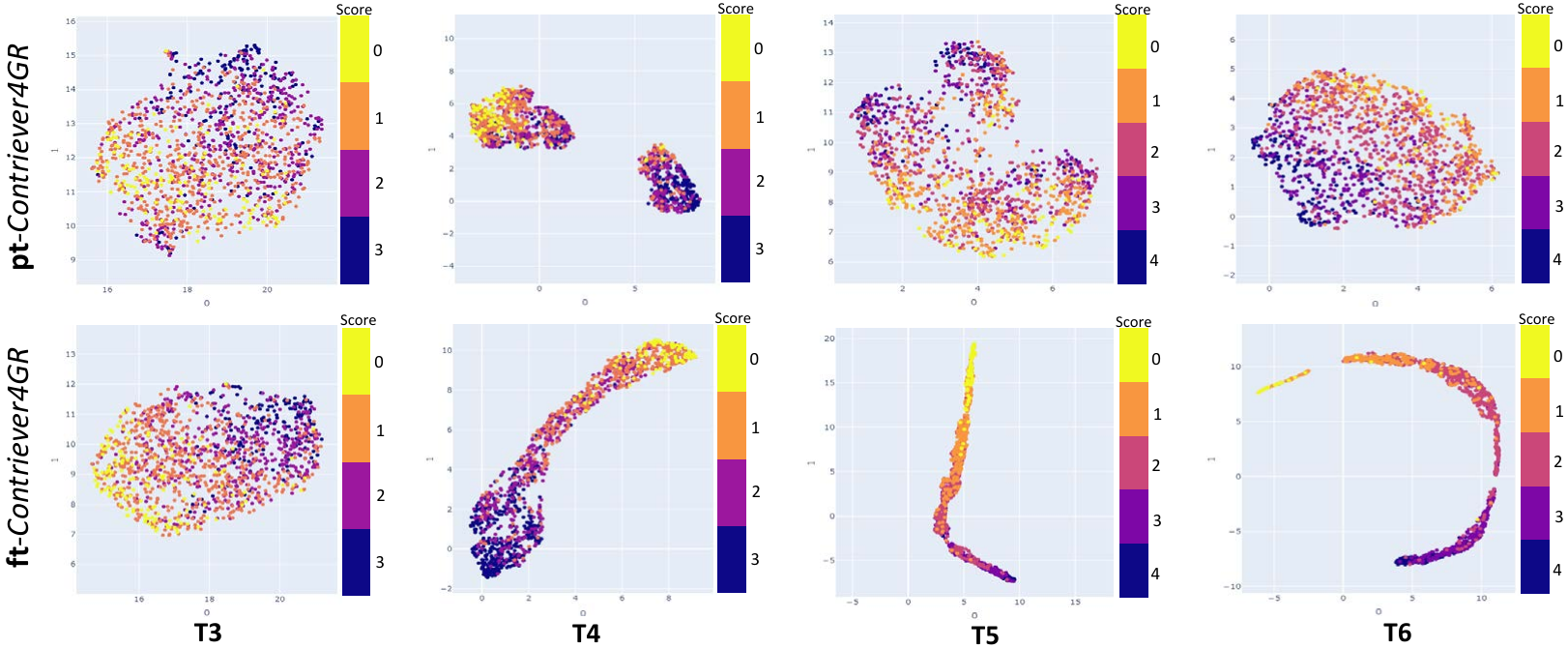}
    \caption{2D Visualization of the training set of each task before and after fine-tuning using UMAP.}
    \label{fig:visualization-TS}
\end{figure*}

\paragraph{\textbf{Visualizing \sys}} In Figure \ref{fig:visualization-TS}, we illustrate the impact of fine-tuning Contriever. These visualizations pertain to the training set of each task. Employing the 2D UMAP dimensional reduction method \cite{mcinnes2018umap}, we plotted the encoded essays in a scatter plot for analysis. The figures at the top show how the pre-trained Contriever encodes the essays. Despite a somewhat mixed arrangement of essays with varying scores, distinct groups corresponding to each relevance level are somewhat discernible. 
However, after fine-tuning, there is a notable refinement, with a more clear ordinal hierarchy that emerges between each relevance level. 
Furthermore, it is observable that the fine-tuning process not only enhances score-level distinctions but also contributes to the compactness of each cluster of essays. This effect is particularly noticeable in tasks 4, 5, and 6, demonstrating the impact of the employed loss function.

\subsection{Few-shot Learning (RQ3)}
\label{sec:RQ3}

\begin{figure*}
    \centering
    \includegraphics[width=\textwidth]{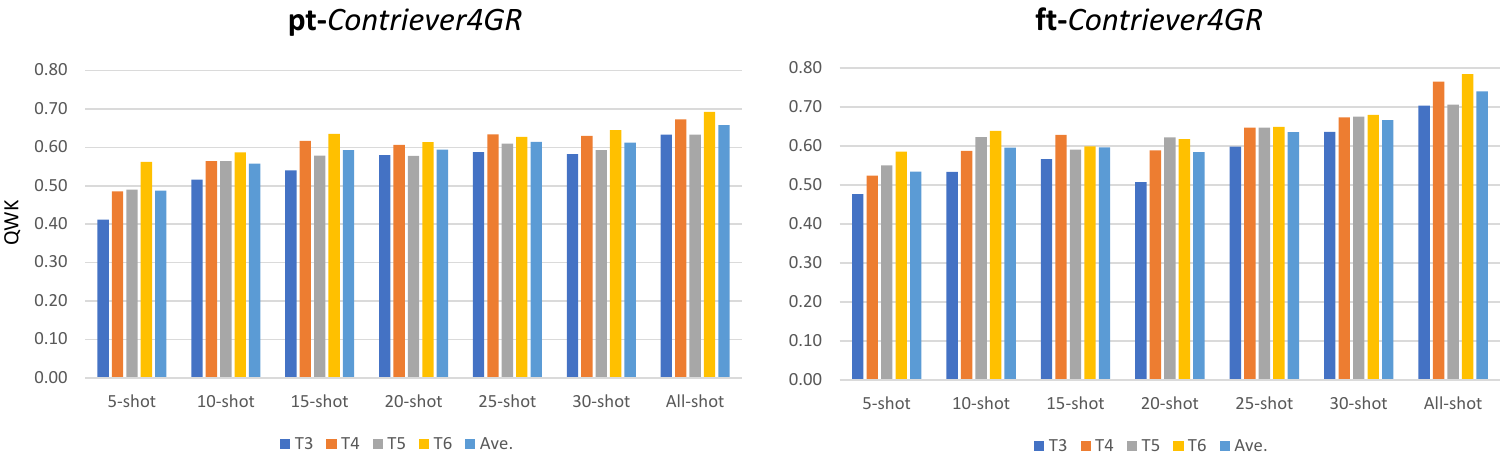}
    \caption{Performance of our models with the $k$-shot setup compared to the full-shot setup on the \underline{test} sets.}
    \label{fig:few-shots}
\end{figure*}

In earlier scenarios, we assume many \emph{labeled} essays from the same task are available (more than a thousand in ASAP++ dataset). While this is a common setup, it is indeed \emph{impractical} in educational contexts and poses a considerable challenge, as it requires time-consuming manual grading for any new task.
\emph{Few-shot} learning alleviates this constraint by requiring only a few graded essays to effectively train the model, making our approach more feasible.

To that end, we study the performance of our pre-trained and fine-tuned models when trained within a $k$-shot setup, where $k$ is the number of labeled essays per relevance level that are available for training for a given task. We varied $k$ from 5 to 30 with a step of 5. 
As the number of shots increases, we augmented the initial set by randomly sampling additional 5 essays per relevance level, creating comparable subsets across experiments.
We then test the trained models on the standard test set of the task.
To improve robustness and mitigate the influence of chance, we repeated the entire process 5 times and reports the average performance over the 5 runs.

Figure \ref{fig:few-shots} illustrates the performance of both \pt{} and \ft{} with the few-shot setup compared to the full-shot setup (i.e., training with \emph{all} labeled examples). 

As anticipated, a general observation reveals that increasing $k$ positively correlates with improved performance, and the fine-tuned models generally outperform their pre-trained counterparts. 

Most notably, the average performance of \ft{} with 30 shots trails by approximately only 7.3 points compared to the full-shot setup; it is crucial to consider that 30 shots imply only 150 labeled essays (assuming 5 relevance levels), which constitutes less than 15\% of the full training set that consists of 1000+ essays, reducing the labeling cost by more than 85\% while sacrificing about 10\% of the performance. This clearly has the potential to control the trade-off between practicality and performance; it indeed opens the door for future efforts toward narrowing the performance gap.

Interestingly, we also noticed that fine-tuning with 25 shots per relevance level was sufficient to reach a similar performance as the pre-trained model when trained with the full set. This highlights the power of the Contriever model when fine-tuned with few shots.

All in all, our models yielded promising performance in the few-shot scenario, though the rule of thumb of having more data for better performance remains. Nevertheless, our models showcase commendable overall performance even with a fraction of the original data, emphasizing the practicality of our approach. 

\subsection{Cross-Task Scoring (RQ4)}
\label{sec:RQ4}
\begin{table}[h]
\caption{Performance of cross-task models on test sets.}
\label{crossprompt_results}

\begin{tabular}{lccccc}
\hline
\textbf{Model} & \textbf{T3} & \textbf{T4} & \textbf{T5} & \textbf{T6} & \textbf{Ave.} \\ \hline
ProTACT \cite{do-etal-2023-prompt} & -  & - & - & - & 0.619 \\
PMAES \cite{chen-li-2023-pmaes} & - & - & - & - & 0.584 \\ \hline

\cpvanilla & 0.435 & 0.556 & 0.194 & 0.260 & 0.361 \\ \hline
\cpipt & 0.541 & 0.551 & 0.526 & 0.441 & 0.515 \\
\cpift & 0.593 & 0.657 & 0.344 & 0.607 & 0.550 \\ \hline
\cpspt & 0.379 & 0.494 & 0.381 & 0.315 & 0.392 \\ 
\cpsft & 0.639 & 0.675 & 0.621 & 0.532 & 0.617 \\ \hline

\end{tabular}
\end{table}

What if we do not have any labeled essays for the test task? Can we leverage labeled essays for other tasks to learn a model that can be used to score essays written for a \emph{new} task for which we do not have any labeled essays? This is exactly the \emph{cross-task} scenario.

For this scenario, the setup goes as follows. For each target (i.e., test) task in the dataset, the 3 other tasks (along with their labeled essays) are used for training. Due to the discrepancy of relevance levels across tasks (T3 and T4 range from 0 to 3, while T5 and T6 range from 0 to 4), centroids for relevance levels 0 to 3 are computed from all of the 3 training tasks, while the centroid for relevance level 4 is computed only from T5 and/or T6.

We experimented with 3 variants of \sys{} for the cross-task scenario. The first is the \emph{vanilla} cross-task model, denoted by the prefix \textbf{ct$^v$}, which directly uses the centroids of the essays' vectors as done for the task-specific scenarios. The second is the \emph{task-independent} version denoted by the prefix \textbf{ct$^i$}, in which we made the essay vectors (thus the centroids) task-independent by semantically ``disentangling'' their corresponding task-prompts. The third is an extended version of the second that is denoted by the prefix \textbf{ct$^s$}, in which we also involve the similarity with the target task-prompt when we score a given test essay. We tried the second and third variants with both \pt{} and \ft{} models. For fine-tuning, we used the same hyper-parameters used in the task-specific experiments but over only one epoch. 

Table \ref{crossprompt_results} presents the performance of the variants of our model compared to the SOTA baselines for the cross-task setup. 
We can draw several observations. First, comparing the performance of the vanilla model with the performance of the task-independent models shows that the idea of excluding the task-prompt representation from the representation of the training essays was very effective, resulting in 15-point and 19-point average improvements in QWK with the pre-trained and fine-tuned models respectively over the vanilla model. Those noticeable improvements indicate that, with this simple trick, the per-task centroids of the same relevance level become indeed better aligned to represent the cluster of their relevance level independent of their corresponding tasks.

Furthermore, the incorporation of the similarity score between the essay and the target task-prompt contributed an additional improvement when applied to the fine-tuned model, achieving a comparable performance to the SOTA model and showing the effectiveness of our approach. This also emphasizes the need to fine-tune the Contriever model for our domain, which results in more distinct clusters for each relevance level. This, in turn, brings the task-prompt vector closer to the cluster of higher-quality essays, consequently improving the similarity score between the essays and their task-prompts.

\section{Conclusion and Future Work} \label{sec:conclusion}

In this study, we introduced a novel approach for graded relevance scoring of written essays that leverages dense retrieval, employing Contriever in particular as an example of unsupervised dense retrieval model. Out of the box, without any fine-tuning, our method yielded promising results, yet post-fine-tuning, it established a new SOTA performance for task-specific scenarios. Furthermore, we proposed a simple adjustment to our approach, eliminating the influence of the task-prompt to enable its adaptability for cross-task settings, achieving a performance that is on par with SOTA baselines. 
Moreover, we showed that our approach exhibited reasonable performance in more practical scenarios where only few essays are labeled for the target writing task. In particular, with only 30 graded essays per relevance level, we observed about 10\% drop in performance while saving more than 85\% of the manual labor cost.
 
However, it is important to acknowledge a limitation of our method, namely its assumption of discrete scores, which may not always hold true. Moreover, in cross-task scenarios, our method does not readily adapt to changes in the score range between source tasks and the target task.

This work opens doors for several future directions. One is to test the ability of other dense retrieval models in our task. We also recognize the need to refine the fine-tuning stage, particularly the loss function, e.g., treating negative examples differently. Finally, we plan to explore ways to better adapt to the score range difference across tasks, and to employ the dense representation of individual training examples in addition to their centroids in scoring.

\begin{acks}
The work of Salam Albatarni was supported by GSRA grant\# GSRA10-L-2-0521-23037, and the work of Sohaila Eltanbouly and Tamer Elsayed was made possible by NPRP grant\# NPRP14S-0402-210127, both from the Qatar National Research Fund (a member of Qatar Foundation). The statements made herein are solely the responsibility of the authors.
\end{acks}

\newpage

\bibliographystyle{ACM-Reference-Format}
\bibliography{sample-base}


\begin{thebibliography}{28}


\ifx \showCODEN    \undefined \def \showCODEN     #1{\unskip}     \fi
\ifx \showDOI      \undefined \def \showDOI       #1{#1}\fi
\ifx \showISBNx    \undefined \def \showISBNx     #1{\unskip}     \fi
\ifx \showISBNxiii \undefined \def \showISBNxiii  #1{\unskip}     \fi
\ifx \showISSN     \undefined \def \showISSN      #1{\unskip}     \fi
\ifx \showLCCN     \undefined \def \showLCCN      #1{\unskip}     \fi
\ifx \shownote     \undefined \def \shownote      #1{#1}          \fi
\ifx \showarticletitle \undefined \def \showarticletitle #1{#1}   \fi
\ifx \showURL      \undefined \def \showURL       {\relax}        \fi
\providecommand\bibfield[2]{#2}
\providecommand\bibinfo[2]{#2}
\providecommand\natexlab[1]{#1}
\providecommand\showeprint[2][]{arXiv:#2}

\bibitem[Cao et~al\mbox{.}(2020)]%
        {cao2020domain}
\bibfield{author}{\bibinfo{person}{Yue Cao}, \bibinfo{person}{Hanqi Jin}, \bibinfo{person}{Xiaojun Wan}, {and} \bibinfo{person}{Zhiwei Yu}.} \bibinfo{year}{2020}\natexlab{}.
\newblock \showarticletitle{Domain-adaptive neural automated essay scoring}. In \bibinfo{booktitle}{\emph{Proceedings of the 43rd International ACM SIGIR Conference on Research and Development in Information Retrieval}}. \bibinfo{pages}{1011--1020}.
\newblock


\bibitem[Chen and Li(2018)]%
        {chen2018relevance}
\bibfield{author}{\bibinfo{person}{Minping Chen} {and} \bibinfo{person}{Xia Li}.} \bibinfo{year}{2018}\natexlab{}.
\newblock \showarticletitle{Relevance-based automated essay scoring via hierarchical recurrent model}. In \bibinfo{booktitle}{\emph{2018 International Conference on Asian Language Processing (IALP)}}. IEEE, \bibinfo{pages}{378--383}.
\newblock


\bibitem[Chen and Li(2023)]%
        {chen-li-2023-pmaes}
\bibfield{author}{\bibinfo{person}{Yuan Chen} {and} \bibinfo{person}{Xia Li}.} \bibinfo{year}{2023}\natexlab{}.
\newblock \showarticletitle{{PMAES}: Prompt-mapping Contrastive Learning for Cross-prompt Automated Essay Scoring}. In \bibinfo{booktitle}{\emph{Proceedings of the 61st Annual Meeting of the Association for Computational Linguistics (Volume 1: Long Papers)}}, \bibfield{editor}{\bibinfo{person}{Anna Rogers}, \bibinfo{person}{Jordan Boyd-Graber}, {and} \bibinfo{person}{Naoaki Okazaki}} (Eds.). \bibinfo{publisher}{Association for Computational Linguistics}, \bibinfo{address}{Toronto, Canada}, \bibinfo{pages}{1489--1503}.
\newblock
\urldef\tempurl%
\url{https://doi.org/10.18653/v1/2023.acl-long.83}
\showDOI{\tempurl}


\bibitem[Cohen(1968)]%
        {cohen1968weighted}
\bibfield{author}{\bibinfo{person}{Jacob Cohen}.} \bibinfo{year}{1968}\natexlab{}.
\newblock \showarticletitle{Weighted kappa: nominal scale agreement provision for scaled disagreement or partial credit.}
\newblock \bibinfo{journal}{\emph{Psychological bulletin}} \bibinfo{volume}{70}, \bibinfo{number}{4} (\bibinfo{year}{1968}), \bibinfo{pages}{213}.
\newblock


\bibitem[Cozma et~al\mbox{.}(2018)]%
        {cozma-etal-2018-automated}
\bibfield{author}{\bibinfo{person}{M{\u{a}}d{\u{a}}lina Cozma}, \bibinfo{person}{Andrei Butnaru}, {and} \bibinfo{person}{Radu~Tudor Ionescu}.} \bibinfo{year}{2018}\natexlab{}.
\newblock \showarticletitle{Automated essay scoring with string kernels and word embeddings}. In \bibinfo{booktitle}{\emph{Proceedings of the 56th Annual Meeting of the Association for Computational Linguistics (Volume 2: Short Papers)}}, \bibfield{editor}{\bibinfo{person}{Iryna Gurevych} {and} \bibinfo{person}{Yusuke Miyao}} (Eds.). \bibinfo{publisher}{Association for Computational Linguistics}, \bibinfo{address}{Melbourne, Australia}, \bibinfo{pages}{503--509}.
\newblock
\urldef\tempurl%
\url{https://doi.org/10.18653/v1/P18-2080}
\showDOI{\tempurl}


\bibitem[Cummins et~al\mbox{.}(2016)]%
        {cummins2016unsupervised}
\bibfield{author}{\bibinfo{person}{Ronan Cummins}, \bibinfo{person}{Helen Yannakoudakis}, {and} \bibinfo{person}{Ted Briscoe}.} \bibinfo{year}{2016}\natexlab{}.
\newblock \showarticletitle{Unsupervised modeling of topical relevance in L2 learner text}. In \bibinfo{booktitle}{\emph{Proceedings of the 11th workshop on innovative use of NLP for building educational applications}}. \bibinfo{pages}{95--104}.
\newblock


\bibitem[Do et~al\mbox{.}(2023)]%
        {do-etal-2023-prompt}
\bibfield{author}{\bibinfo{person}{Heejin Do}, \bibinfo{person}{Yunsu Kim}, {and} \bibinfo{person}{Gary~Geunbae Lee}.} \bibinfo{year}{2023}\natexlab{}.
\newblock \showarticletitle{Prompt- and Trait Relation-aware Cross-prompt Essay Trait Scoring}. In \bibinfo{booktitle}{\emph{Findings of the Association for Computational Linguistics: ACL 2023}}, \bibfield{editor}{\bibinfo{person}{Anna Rogers}, \bibinfo{person}{Jordan Boyd-Graber}, {and} \bibinfo{person}{Naoaki Okazaki}} (Eds.). \bibinfo{publisher}{Association for Computational Linguistics}, \bibinfo{address}{Toronto, Canada}, \bibinfo{pages}{1538--1551}.
\newblock
\urldef\tempurl%
\url{https://doi.org/10.18653/v1/2023.findings-acl.98}
\showDOI{\tempurl}


\bibitem[Dong et~al\mbox{.}(2017)]%
        {dong2017attention}
\bibfield{author}{\bibinfo{person}{Fei Dong}, \bibinfo{person}{Yue Zhang}, {and} \bibinfo{person}{Jie Yang}.} \bibinfo{year}{2017}\natexlab{}.
\newblock \showarticletitle{Attention-based Recurrent Convolutional Neural Network for Automatic Essay Scoring}. In \bibinfo{booktitle}{\emph{Proceedings of the 21st Conference on Computational Natural Language Learning ({C}o{NLL} 2017)}}, \bibfield{editor}{\bibinfo{person}{Roger Levy} {and} \bibinfo{person}{Lucia Specia}} (Eds.). \bibinfo{publisher}{Association for Computational Linguistics}, \bibinfo{address}{Vancouver, Canada}, \bibinfo{pages}{153--162}.
\newblock
\urldef\tempurl%
\url{https://doi.org/10.18653/v1/K17-1017}
\showDOI{\tempurl}


\bibitem[Gao et~al\mbox{.}(2023)]%
        {gao-etal-2023-precise}
\bibfield{author}{\bibinfo{person}{Luyu Gao}, \bibinfo{person}{Xueguang Ma}, \bibinfo{person}{Jimmy Lin}, {and} \bibinfo{person}{Jamie Callan}.} \bibinfo{year}{2023}\natexlab{}.
\newblock \showarticletitle{Precise Zero-Shot Dense Retrieval without Relevance Labels}. In \bibinfo{booktitle}{\emph{Proceedings of the 61st Annual Meeting of the Association for Computational Linguistics (Volume 1: Long Papers)}}, \bibfield{editor}{\bibinfo{person}{Anna Rogers}, \bibinfo{person}{Jordan Boyd-Graber}, {and} \bibinfo{person}{Naoaki Okazaki}} (Eds.). \bibinfo{publisher}{Association for Computational Linguistics}, \bibinfo{address}{Toronto, Canada}, \bibinfo{pages}{1762--1777}.
\newblock
\urldef\tempurl%
\url{https://doi.org/10.18653/v1/2023.acl-long.99}
\showDOI{\tempurl}


\bibitem[Izacard et~al\mbox{.}(2021)]%
        {izacard2021towards}
\bibfield{author}{\bibinfo{person}{Gautier Izacard}, \bibinfo{person}{Mathilde Caron}, \bibinfo{person}{Lucas Hosseini}, \bibinfo{person}{Sebastian Riedel}, \bibinfo{person}{Piotr Bojanowski}, \bibinfo{person}{Armand Joulin}, {and} \bibinfo{person}{Edouard Grave}.} \bibinfo{year}{2021}\natexlab{}.
\newblock \showarticletitle{Unsupervised dense information retrieval with contrastive learning}.
\newblock \bibinfo{journal}{\emph{arXiv preprint arXiv:2112.09118}} (\bibinfo{year}{2021}).
\newblock


\bibitem[Jiang et~al\mbox{.}(2023)]%
        {jiang-etal}
\bibfield{author}{\bibinfo{person}{Zhiwei Jiang}, \bibinfo{person}{Tianyi Gao}, \bibinfo{person}{Yafeng Yin}, \bibinfo{person}{Meng Liu}, \bibinfo{person}{Hua Yu}, \bibinfo{person}{Zifeng Cheng}, {and} \bibinfo{person}{Qing Gu}.} \bibinfo{year}{2023}\natexlab{}.
\newblock \showarticletitle{Improving Domain Generalization for Prompt-Aware Essay Scoring via Disentangled Representation Learning}. In \bibinfo{booktitle}{\emph{Proceedings of the 61st Annual Meeting of the Association for Computational Linguistics (Volume 1: Long Papers)}}, \bibfield{editor}{\bibinfo{person}{Anna Rogers}, \bibinfo{person}{Jordan Boyd-Graber}, {and} \bibinfo{person}{Naoaki Okazaki}} (Eds.). \bibinfo{publisher}{Association for Computational Linguistics}, \bibinfo{address}{Toronto, Canada}, \bibinfo{pages}{12456--12470}.
\newblock
\urldef\tempurl%
\url{https://doi.org/10.18653/v1/2023.acl-long.696}
\showDOI{\tempurl}


\bibitem[Jin et~al\mbox{.}(2018)]%
        {jin-etal-2018-tdnn}
\bibfield{author}{\bibinfo{person}{Cancan Jin}, \bibinfo{person}{Ben He}, \bibinfo{person}{Kai Hui}, {and} \bibinfo{person}{Le Sun}.} \bibinfo{year}{2018}\natexlab{}.
\newblock \showarticletitle{{TDNN}: A Two-stage Deep Neural Network for Prompt-independent Automated Essay Scoring}. In \bibinfo{booktitle}{\emph{Proceedings of the 56th Annual Meeting of the Association for Computational Linguistics (Volume 1: Long Papers)}}, \bibfield{editor}{\bibinfo{person}{Iryna Gurevych} {and} \bibinfo{person}{Yusuke Miyao}} (Eds.). \bibinfo{publisher}{Association for Computational Linguistics}, \bibinfo{address}{Melbourne, Australia}, \bibinfo{pages}{1088--1097}.
\newblock
\urldef\tempurl%
\url{https://doi.org/10.18653/v1/P18-1100}
\showDOI{\tempurl}


\bibitem[Ke and Ng(2019)]%
        {ke2019automated}
\bibfield{author}{\bibinfo{person}{Zixuan Ke} {and} \bibinfo{person}{Vincent Ng}.} \bibinfo{year}{2019}\natexlab{}.
\newblock \showarticletitle{Automated Essay Scoring: A Survey of the State of the Art.}. In \bibinfo{booktitle}{\emph{IJCAI}}, Vol.~\bibinfo{volume}{19}. \bibinfo{pages}{6300--6308}.
\newblock


\bibitem[Khattab and Zaharia(2020)]%
        {khattab2020colbert}
\bibfield{author}{\bibinfo{person}{Omar Khattab} {and} \bibinfo{person}{Matei Zaharia}.} \bibinfo{year}{2020}\natexlab{}.
\newblock \showarticletitle{Colbert: Efficient and effective passage search via contextualized late interaction over bert}. In \bibinfo{booktitle}{\emph{Proceedings of the 43rd International ACM SIGIR conference on research and development in Information Retrieval}}. \bibinfo{pages}{39--48}.
\newblock


\bibitem[Kumar et~al\mbox{.}(2022)]%
        {kumar2021many}
\bibfield{author}{\bibinfo{person}{Rahul Kumar}, \bibinfo{person}{Sandeep Mathias}, \bibinfo{person}{Sriparna Saha}, {and} \bibinfo{person}{Pushpak Bhattacharyya}.} \bibinfo{year}{2022}\natexlab{}.
\newblock \showarticletitle{Many Hands Make Light Work: Using Essay Traits to Automatically Score Essays}. In \bibinfo{booktitle}{\emph{Proceedings of the 2022 Conference of the North American Chapter of the Association for Computational Linguistics: Human Language Technologies}}, \bibfield{editor}{\bibinfo{person}{Marine Carpuat}, \bibinfo{person}{Marie-Catherine de~Marneffe}, {and} \bibinfo{person}{Ivan~Vladimir Meza~Ruiz}} (Eds.). \bibinfo{publisher}{Association for Computational Linguistics}, \bibinfo{address}{Seattle, United States}, \bibinfo{pages}{1485--1495}.
\newblock
\urldef\tempurl%
\url{https://doi.org/10.18653/v1/2022.naacl-main.106}
\showDOI{\tempurl}


\bibitem[Li et~al\mbox{.}(2020)]%
        {li2020sednn}
\bibfield{author}{\bibinfo{person}{Xia Li}, \bibinfo{person}{Minping Chen}, {and} \bibinfo{person}{Jian-Yun Nie}.} \bibinfo{year}{2020}\natexlab{}.
\newblock \showarticletitle{SEDNN: Shared and enhanced deep neural network model for cross-prompt automated essay scoring}.
\newblock \bibinfo{journal}{\emph{Knowledge-Based Systems}}  \bibinfo{volume}{210} (\bibinfo{year}{2020}), \bibinfo{pages}{106491}.
\newblock


\bibitem[Mathias and Bhattacharyya(2018)]%
        {mathias2018asap++}
\bibfield{author}{\bibinfo{person}{Sandeep Mathias} {and} \bibinfo{person}{Pushpak Bhattacharyya}.} \bibinfo{year}{2018}\natexlab{}.
\newblock \showarticletitle{{ASAP}++: Enriching the {ASAP} Automated Essay Grading Dataset with Essay Attribute Scores}. In \bibinfo{booktitle}{\emph{Proceedings of the Eleventh International Conference on Language Resources and Evaluation ({LREC} 2018)}}, \bibfield{editor}{\bibinfo{person}{Nicoletta Calzolari}, \bibinfo{person}{Khalid Choukri}, \bibinfo{person}{Christopher Cieri}, \bibinfo{person}{Thierry Declerck}, \bibinfo{person}{Sara Goggi}, \bibinfo{person}{Koiti Hasida}, \bibinfo{person}{Hitoshi Isahara}, \bibinfo{person}{Bente Maegaard}, \bibinfo{person}{Joseph Mariani}, \bibinfo{person}{H{\'e}l{\`e}ne Mazo}, \bibinfo{person}{Asuncion Moreno}, \bibinfo{person}{Jan Odijk}, \bibinfo{person}{Stelios Piperidis}, {and} \bibinfo{person}{Takenobu Tokunaga}} (Eds.). \bibinfo{publisher}{European Language Resources Association (ELRA)}, \bibinfo{address}{Miyazaki, Japan}.
\newblock
\urldef\tempurl%
\url{https://aclanthology.org/L18-1187}
\showURL{%
\tempurl}


\bibitem[Mathias and Bhattacharyya(2020)]%
        {mathias-bhattacharyya-2020-neural}
\bibfield{author}{\bibinfo{person}{Sandeep Mathias} {and} \bibinfo{person}{Pushpak Bhattacharyya}.} \bibinfo{year}{2020}\natexlab{}.
\newblock \showarticletitle{Can Neural Networks Automatically Score Essay Traits?}. In \bibinfo{booktitle}{\emph{Proceedings of the Fifteenth Workshop on Innovative Use of NLP for Building Educational Applications}}, \bibfield{editor}{\bibinfo{person}{Jill Burstein}, \bibinfo{person}{Ekaterina Kochmar}, \bibinfo{person}{Claudia Leacock}, \bibinfo{person}{Nitin Madnani}, \bibinfo{person}{Ildik{\'o} Pil{\'a}n}, \bibinfo{person}{Helen Yannakoudakis}, {and} \bibinfo{person}{Torsten Zesch}} (Eds.). \bibinfo{publisher}{Association for Computational Linguistics}, \bibinfo{address}{Seattle, WA, USA → Online}, \bibinfo{pages}{85--91}.
\newblock
\urldef\tempurl%
\url{https://doi.org/10.18653/v1/2020.bea-1.8}
\showDOI{\tempurl}


\bibitem[McInnes et~al\mbox{.}(2018)]%
        {mcinnes2018umap}
\bibfield{author}{\bibinfo{person}{Leland McInnes}, \bibinfo{person}{John Healy}, {and} \bibinfo{person}{James Melville}.} \bibinfo{year}{2018}\natexlab{}.
\newblock \showarticletitle{Umap: Uniform manifold approximation and projection for dimension reduction}.
\newblock \bibinfo{journal}{\emph{arXiv preprint arXiv:1802.03426}} (\bibinfo{year}{2018}).
\newblock


\bibitem[Oord et~al\mbox{.}(2018)]%
        {oord2018representation}
\bibfield{author}{\bibinfo{person}{Aaron van~den Oord}, \bibinfo{person}{Yazhe Li}, {and} \bibinfo{person}{Oriol Vinyals}.} \bibinfo{year}{2018}\natexlab{}.
\newblock \showarticletitle{Representation learning with contrastive predictive coding}.
\newblock \bibinfo{journal}{\emph{arXiv preprint arXiv:1807.03748}} (\bibinfo{year}{2018}).
\newblock


\bibitem[Persing and Ng(2014)]%
        {persing-modeling}
\bibfield{author}{\bibinfo{person}{Isaac Persing} {and} \bibinfo{person}{Vincent Ng}.} \bibinfo{year}{2014}\natexlab{}.
\newblock \showarticletitle{Modeling Prompt Adherence in Student Essays}. In \bibinfo{booktitle}{\emph{Proceedings of the 52nd Annual Meeting of the Association for Computational Linguistics (Volume 1: Long Papers)}}, \bibfield{editor}{\bibinfo{person}{Kristina Toutanova} {and} \bibinfo{person}{Hua Wu}} (Eds.). \bibinfo{publisher}{Association for Computational Linguistics}, \bibinfo{address}{Baltimore, Maryland}, \bibinfo{pages}{1534--1543}.
\newblock
\urldef\tempurl%
\url{https://doi.org/10.3115/v1/P14-1144}
\showDOI{\tempurl}


\bibitem[Rei and Cummins(2016)]%
        {rei-cummins-2016-sentence}
\bibfield{author}{\bibinfo{person}{Marek Rei} {and} \bibinfo{person}{Ronan Cummins}.} \bibinfo{year}{2016}\natexlab{}.
\newblock \showarticletitle{Sentence Similarity Measures for Fine-Grained Estimation of Topical Relevance in Learner Essays}. In \bibinfo{booktitle}{\emph{Proceedings of the 11th Workshop on Innovative Use of {NLP} for Building Educational Applications}}, \bibfield{editor}{\bibinfo{person}{Joel Tetreault}, \bibinfo{person}{Jill Burstein}, \bibinfo{person}{Claudia Leacock}, {and} \bibinfo{person}{Helen Yannakoudakis}} (Eds.). \bibinfo{publisher}{Association for Computational Linguistics}, \bibinfo{address}{San Diego, CA}, \bibinfo{pages}{283--288}.
\newblock
\urldef\tempurl%
\url{https://doi.org/10.18653/v1/W16-0533}
\showDOI{\tempurl}


\bibitem[Ridley et~al\mbox{.}(2020)]%
        {ridley2020prompt}
\bibfield{author}{\bibinfo{person}{Robert Ridley}, \bibinfo{person}{Liang He}, \bibinfo{person}{Xinyu Dai}, \bibinfo{person}{Shujian Huang}, {and} \bibinfo{person}{Jiajun Chen}.} \bibinfo{year}{2020}\natexlab{}.
\newblock \showarticletitle{Prompt agnostic essay scorer: a domain generalization approach to cross-prompt automated essay scoring}.
\newblock \bibinfo{journal}{\emph{arXiv preprint arXiv:2008.01441}} (\bibinfo{year}{2020}).
\newblock


\bibitem[Ridley et~al\mbox{.}(2021)]%
        {ridley2021automated}
\bibfield{author}{\bibinfo{person}{Robert Ridley}, \bibinfo{person}{Liang He}, \bibinfo{person}{Xin-yu Dai}, \bibinfo{person}{Shujian Huang}, {and} \bibinfo{person}{Jiajun Chen}.} \bibinfo{year}{2021}\natexlab{}.
\newblock \showarticletitle{Automated cross-prompt scoring of essay traits}. In \bibinfo{booktitle}{\emph{Proceedings of the AAAI conference on artificial intelligence}}, Vol.~\bibinfo{volume}{35}. \bibinfo{pages}{13745--13753}.
\newblock


\bibitem[Song et~al\mbox{.}(2020)]%
        {ijcai2020p536}
\bibfield{author}{\bibinfo{person}{Wei Song}, \bibinfo{person}{Ziyao Song}, \bibinfo{person}{Lizhen Liu}, {and} \bibinfo{person}{Ruiji Fu}.} \bibinfo{year}{2020}\natexlab{}.
\newblock \showarticletitle{Hierarchical Multi-task Learning for Organization Evaluation of Argumentative Student Essays}. In \bibinfo{booktitle}{\emph{Proceedings of the Twenty-Ninth International Joint Conference on Artificial Intelligence, {IJCAI-20}}}, \bibfield{editor}{\bibinfo{person}{Christian Bessiere}} (Ed.). \bibinfo{publisher}{International Joint Conferences on Artificial Intelligence Organization}, \bibinfo{pages}{3875--3881}.
\newblock
\urldef\tempurl%
\url{https://doi.org/10.24963/ijcai.2020/536}
\showDOI{\tempurl}
\newblock
\shownote{Main track}.


\bibitem[Taghipour and Ng(2016)]%
        {taghipourng2016neural}
\bibfield{author}{\bibinfo{person}{Kaveh Taghipour} {and} \bibinfo{person}{Hwee~Tou Ng}.} \bibinfo{year}{2016}\natexlab{}.
\newblock \showarticletitle{A Neural Approach to Automated Essay Scoring}. In \bibinfo{booktitle}{\emph{Proceedings of the 2016 Conference on Empirical Methods in Natural Language Processing}}, \bibfield{editor}{\bibinfo{person}{Jian Su}, \bibinfo{person}{Kevin Duh}, {and} \bibinfo{person}{Xavier Carreras}} (Eds.). \bibinfo{publisher}{Association for Computational Linguistics}, \bibinfo{address}{Austin, Texas}, \bibinfo{pages}{1882--1891}.
\newblock
\urldef\tempurl%
\url{https://doi.org/10.18653/v1/D16-1193}
\showDOI{\tempurl}


\bibitem[Xue et~al\mbox{.}(2021)]%
        {9530411}
\bibfield{author}{\bibinfo{person}{Jin Xue}, \bibinfo{person}{Xiaoyi Tang}, {and} \bibinfo{person}{Liyan Zheng}.} \bibinfo{year}{2021}\natexlab{}.
\newblock \showarticletitle{A Hierarchical BERT-Based Transfer Learning Approach for Multi-Dimensional Essay Scoring}.
\newblock \bibinfo{journal}{\emph{IEEE Access}}  \bibinfo{volume}{9} (\bibinfo{year}{2021}), \bibinfo{pages}{125403--125415}.
\newblock
\urldef\tempurl%
\url{https://doi.org/10.1109/ACCESS.2021.3110683}
\showDOI{\tempurl}


\bibitem[Yang et~al\mbox{.}(2020)]%
        {yang-etal-2020-enhancing}
\bibfield{author}{\bibinfo{person}{Ruosong Yang}, \bibinfo{person}{Jiannong Cao}, \bibinfo{person}{Zhiyuan Wen}, \bibinfo{person}{Youzheng Wu}, {and} \bibinfo{person}{Xiaodong He}.} \bibinfo{year}{2020}\natexlab{}.
\newblock \showarticletitle{Enhancing Automated Essay Scoring Performance via Fine-tuning Pre-trained Language Models with Combination of Regression and Ranking}. In \bibinfo{booktitle}{\emph{Findings of the Association for Computational Linguistics: EMNLP 2020}}, \bibfield{editor}{\bibinfo{person}{Trevor Cohn}, \bibinfo{person}{Yulan He}, {and} \bibinfo{person}{Yang Liu}} (Eds.). \bibinfo{publisher}{Association for Computational Linguistics}, \bibinfo{address}{Online}, \bibinfo{pages}{1560--1569}.
\newblock
\urldef\tempurl%
\url{https://doi.org/10.18653/v1/2020.findings-emnlp.141}
\showDOI{\tempurl}


\end{thebibliography}

\end{document}